\documentclass[superscriptaddress,aps,prb,twocolumn]{revtex4}
\usepackage{bm}
\usepackage{graphics}
\usepackage{graphicx}
\usepackage[dvipsnames]{xcolor}
\usepackage[colorlinks=true,linkcolor=blue,citecolor=BrickRed]{hyperref}
\begin{document}

\title{Theory of disordered unconventional superconductors }

\author {A. Keles}

\affiliation{Department of Physics, University of Washington, Seattle, WA 98195}

\author{A. V. Andreev}

\affiliation{Department of Physics, University of Washington, Seattle, WA 98195}

\author {S. A. Kivelson}

\affiliation{Department of Physics, Stanford University, CA 94305}

\author {B. Z. Spivak}

\affiliation{Department of Physics, University of Washington, Seattle, WA 98195}

\begin{abstract}

\emph{This article is dedicated to A. F. Andreev on the occasion of his 75th birthday.}\\

In contrast to conventional s-wave superconductivity, unconventional (e.g. p or
d-wave) superconductivity is strongly suppressed even by relatively weak
disorder.  Upon approaching the superconductor-metal transition, the order
parameter amplitude becomes increasingly inhomogeneous leading to effective
granularity and a phase ordering transition described by the Mattis model of
spin glasses.  One consequence of this is that at low enough temperatures,
between the clean unconventional superconducting and the diffusive metallic
phases,  there is necessarily an  intermediate superconducting phase which
exhibits s-wave symmetry on macroscopic scales.

\end{abstract}
\date{\today}

\maketitle

\section{Introduction}

Generally, the  superconducting order parameter depends on two coordinates and
two spin indices, $\Delta_{\alpha\beta}({\bf r}, {\bf r}')$.  A classification
of possible superconducting phases in crystalline materials was given in
Refs.~\onlinecite{VolovikGork} and \onlinecite{sigristueda}. The majority of
crystalline superconductors with low transition temperatures have a singlet order
parameter with an s-wave orbital symmetry that does not change under
rotation of the coordinates. In the simplest case, 
$\Delta_{\alpha\beta}({\bf r},{\bf r}')
\approx i
(\hat\sigma_2)_{\alpha\beta}\delta({\bf r}-{\bf r}')\Delta^{(s)}({\bf r})$
depends significantly only on a single coordinate, where
$\hat\sigma_2$ is the second Pauli matrix in spin space, $\Delta^{(s)}({\bf r})$ is
a complex valued function and the superscript $s$ indicates that is has s-wave
symmetry.  However, over the last decades a number of superconductors have been
discovered in which the order parameter transforms according to a non-trivial
representation of the point group of the underlying crystal.  Although such
superconductors are quite common by now, following the terminology of
Ref.~\onlinecite{mineev}, we refer to them as being ``unconventional.''

Important examples include the high temperature cuprate superconductors which
have a singlet d-wave symmetry\cite{sigristueda,tsuei}:
$\Delta_{\alpha\beta}({\bf r}, {\bf r}')= i
(\hat\sigma_2)_{\alpha\beta}\Delta^{(d)}({\bf r}- {\bf r}')$ where
$\Delta^{(d)}({\bf r}- {\bf r}')$ changes sign under coordinate rotation by
$\pi/2$.  The best known example of a p-wave superfluid is superfluid $^3{\rm
	He}$.   One of the leading candidates for p-wave pairing in electronic systems
is Sr$_{2}$RuO$_{4}$.\cite{rmp_pwave} There are numerous pieces of experimental
evidence that the  superconducting state of Sr$_{2}$RuO$_{4}$ has odd parity,
breaks time reversal symmetry and is spin
triplet.\cite{nelson,kidwingira,Xia,Luke,rmp_pwave,maeno,caveat} An order parameter
consistent with these experiments is given by the chiral p-wave state
\cite{sigrist} which has the form $\Delta_{\alpha\beta}( {\bf p}) \sim p_{x}\pm
ip_{y}$ where $\Delta_{\alpha\beta} ({\bf p})$ is the Fourier transform of
$\Delta_{\alpha\beta}({\bf r}-{\bf r}')$.  Anderson's theorem accounts for the
fact that superconductivity in s-wave superconductors is destroyed only when the
disorder is so strong that $p_{F}l\sim 1$, where $p_{F}$ is the Fermi momentum
and $l$ is the electronic elastic mean free path.  However, since in
unconventional superconductors $\Delta_{\alpha\beta} ({\bf p})$ depends on the
direction of the relative momentum ${\bf p}$ of electrons in the Cooper pair,
they are much more sensitive to disorder; even at temperature $T = 0$,
unconventional  superconductivity is destroyed when $l$ is comparable to the
zero temperature coherence length, $\xi_0$, in the pure superconductor, $l\sim \xi_{0}\gg
1/p_{F}$.  The fate of  unconventional superconductivity subject to increasing
disorder depends on the sign of the interaction constant in the s-wave channel.
It is straightforward to see that if the interaction in the s-wave channel is
attractive, but weaker than the attraction in an unconventional channel, then as
a function of increasing disorder, there will first be a transition from the
unconventional to an s-wave phase when $l \sim \xi_0$, followed  by a transition
to a non-superconducting phase when $l \sim p_F^{-1}$.

In this article we consider the more interesting and realistic case, in which
the interaction in the s-channel is repulsive.  In this case, we show that there
is necessarily a range of disorder strength in which, although locally the
pairing remains unconventional, the system has a global s-wave symmetry with
respect to any macroscopic superconducting interference experiments.  Therefore
there must be at least two phase transitions as a function of increasing
disorder: a d-wave (or p-wave) to s-wave, followed by an s-wave to normal metal
transition. Qualitatively the phase diagram of disordered unconventional
superconductors is shown in Fig.~\ref{fig:phase_diagram} (An incomplete
derivation of these results -- only for the d-wave case -- was obtained in
Refs.~\onlinecite{Kivelson} and \onlinecite{Kivelson1}.)

The existence of the intermediate s-wave superconducting phase between the
unconventional superconductor and the normal metal (and of the associated
s-wave to unconventional superconductor transition) can be understood at a mean
field level which neglects both classical and quantum fluctuations of the order
parameter. The electron mean free path is an average characteristic of disorder.
Let us introduce a local value of the mean free path $\bar{l}({\bf r})$ averaged
over regions with a size of order $\xi_{0}$.  When the disorder is sufficiently
strong such that, on average, $\bar{l} < \xi_0$, the superconducting order
parameter will only be large in the rare regions which satisfies $\bar{l}({\bf
	r}) > \xi_0$. In this case, the system can be visualized as a matrix of
superconducting islands that are coupled through  Josephson links in a
non-superconducting metal. (The superconductivity inside an island can also be
enhanced if the pairing interaction is stronger than average, {\it i.e.} if the
local value of $\xi_0$ is anomalously small.) At sufficiently large values of
disorder the distance between the islands is larger than both their size and the
mean free path.

\section{Mattis model description of  disordered  unconventional superconductors}
\label{sec:Mattis_model}

Below we show that in the vicinity of the superconductor-normal metal transition, the
superconducting phase may be described by the Mattis model.

\subsection{An isolated superconducting island}

To begin with, we consider the  mean-field description of an isolated
superconducting island. The order parameter in an individual island is written as
$\hat\Delta_a({\bf r},{\bf r}')$ where hat  indicates the two by two matrix
structure in spin space and we label individual islands with Latin indices $a$,
$b$, $\ldots$.  Generally, %
as a consequence of the random disorder,
neither  the shape of the island, nor the texture of
pairing tendencies within it have any particular symmetry, so the resulting gap
function $\hat\Delta_a({\bf r},{\bf r}^\prime)$ mixes the symmetries of
different bulk phases. Since there is no translational symmetry, it is
convenient to define $\hat\Delta_a(\tilde{\bf  r},{\bf p})$ as the Fourier
transform of $\hat\Delta_a({\bf r},{\bf r}')$ with respect to the relative
coordinate ${\bf r}-{\bf r}^\prime$ and to use 
$\tilde{\bf r}=({\bf
	r}+{\bf r}^\prime)/2$ as the center of mass coordinate. (Since all of the coordinates
to appear from now on will be the center of mass coordinates, we will henceforth drop the tilde. )
In the absence of
spin-orbit coupling, a sharp distinction exists between spin-0 (singlet) and
spin-1 (triplet) pairing, although even that distinction is entirely lost in the
presence of spin-orbit coupling.  The most general form of the gap function
(with a phase convention which we will specify later) expressed as a second rank
spinor in terms of Pauli matrices is
\begin{equation} \label{eq:Delta_Pauli}
    \hat{\Delta}_a(\mathbf{r},\mathbf{p})=
		e^{i \phi_a}\  i\hat \sigma_2 \left(\Delta_a \hat
    1 + \boldsymbol{\Delta}_a \cdot \hat{\boldsymbol{\sigma}} \right).
\end{equation}
where we have left implicit  the $\mathbf{r}$ and $\mathbf{p}$ dependence
of the scalar   $\Delta_a$ and vector $\mathbf{\Delta}_a$ quantities that
represent the singlet and triplet components of the order parameter.

The energy of a single grain is independent of the overall phase of the order
parameter $\phi_a$.  In the absence of spin-orbit interaction it is also
independent of the direction $\mathbf{\Delta}_a$.  An additional discrete
degeneracy may be associated with time-reversal invariance of the problem. The
latter implies that the state described by a time-reversed order parameter
\begin{equation}\label{eq:TR}
	\hat{\bar\Delta}_a(\mathbf{r},\mathbf{p})
	\equiv
		-i \hat\sigma_2
		[\hat{\Delta}_a(\mathbf{r},-\mathbf{p})]^*
		i\hat\sigma_2
	\end{equation}
leads to the same energy of the grain. In the absence of spontaneous breaking of
time reversal symmetry the time reversal operation leads to the same physical
state $\hat{\bar \Delta}_{a}=\hat{\Delta}_{a}$, otherwise the time-reversed
state is physically different.

It is important to note that generally 
(at the present mean-field level)
time reversal symmetry is violated in
droplets of unconventional superconductors of a random shape. This occurs even
in the case 
when the bulk  phase of the unconventional superconductor is time
reversal invariant, such as d-wave superconductors or the $p_{x}$ and $p_{y}$ phases
realized in strained Sr$_{2}$RuO$_4$ \cite{Mackenzie}. For example, d-wave
superconducting droplets of a random shape embedded into a bulk metal
can have, with non-vanishing probability,  a
local geometry analogous to that of a corner SQUID experiment \cite{tsuei}
in which two sides of a droplet with different signs of the order parameter are
connected by a metallic Josephson link with an effective negative critical current.  
An equilibrium
current will thus flow, 
provided the critical
current of the ``negative link" is 
large enough.

We will characterize the degeneracy with respect to time reversal by a
pseudo-spin index $\xi_a=\pm 1$. In this case it is convenient to introduce a
pseudospin $\xi_a$ in each grain that will distinguish the two time-reversed states,
\begin{equation}\label{eq:Delta_a_s}
    \hat{\Delta}_{a}^{\xi_a}(\mathbf{r}, \mathbf{p})=\left\{
    \begin{array}{cc}
      \hat{\Delta}_a(\mathbf{r}, \mathbf{p}), & \xi_a=+1 \\
      \hat{\bar\Delta}_a(\mathbf{r}, \mathbf{p}), & \xi_a=-1
    \end{array}
    \right.
\end{equation}
and write the general expression for the order parameter in each grain as
\begin{equation}
    e^{i \phi_a} \hat{\Delta}^{\xi_a}_{a}(\mathbf{r}, \mathbf{p}) .
    \label{eq:Delta_a_general}
\end{equation}
where we explicitly separate the U(1) phase of the order parameter.

\subsection{Josephson coupling between islands}

Electrons propagating in   non-superconducting metals experience Andreev
reflection~\cite{Andreev} from the superconducting islands. This induces
Josephson coupling between the islands.
So long as the separation between 
islands is large, 
the spatial dependence of the
order parameter within each grain $\hat{\Delta}_a^{\xi_a}(\mathbf{r}, \mathbf{p})$, will not be affected.
Therefore the low energy Hamiltonian of the system may be expressed in terms of
the phases $\phi_a$ only. The energy of this coupling can be expressed in  the
form
\begin{equation}\label{eq:JosEn}
E_{J}=- \frac{1}{2}\sum_{a\neq b} J_{ab}({\xi_a\xi_b})\cos[\phi_{a}-\phi_{b}+\theta_{ab}(\xi_a,\xi_b)]
\end{equation}
Here 
$J_{ab}(\pm 1)>0$  is the Josephson coupling energy between the islands
$a$ and $b$, and  $\theta
_{ab}(\xi_a,\xi_b)$  is the  phase which is determined by the
spatial dependence of the complex order parameter in the grains,
$\hat{\Delta}_a^{\xi_a}(\mathbf{r}, \mathbf{p})$
 (which in turn still depends on which state, $\xi_a=\pm 1$, is involved).

Our goal is to show that 
in the limit in which the distance between the islands is 
sufficiently large compared to
their size then the link phases may be written as
\begin{equation}\label{eq:theta_Mattis}
\theta
_{ab}(\xi_a,\xi_b)
\approx \theta^{\xi_a}_{a}-\theta^{\xi_b}_{b}.
\end{equation}
Eqs.~(\ref{eq:JosEn}), (\ref{eq:theta_Mattis}) represent the XY  Mattis model,
which is well known in the theory of spin glasses\cite{mattis}.  One can gauge
away $\theta_{a}$ reducing Eq.~(\ref{eq:JosEn})  to a conventional form familiar
from the s-wave superconductor, or XY ferromagnet, 

\begin{equation}\label{eq:JosEnMattRed}
	E_{J}
		=-\frac{1}{2}\sum_{a\neq b} E_{ab}^{\xi_a\xi_b}
		=-\frac{1}{2}\sum_{a\neq b} J_{ab}
		({\xi_a\xi_b})\cos\left[\tilde{\phi}
		_{a}-\tilde{\phi}
		_{b}\right]
\end{equation}
where $\tilde{\phi}
_{a}=\phi_{a}+\theta^{\xi_a}_{a}$. Therefore the system is not a
superconducting glass because its ground state has a hidden symmetry.

\begin{figure}[ptb]
\includegraphics[width=0.5\textwidth]{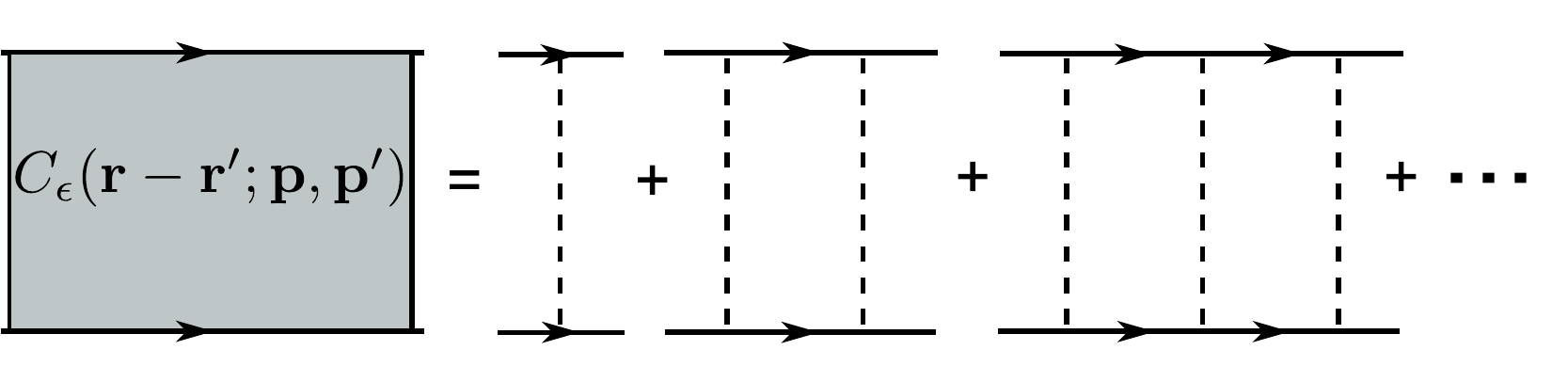}
\caption{Diagrammatic representation of the Cooperon ladder. Solid lines are
	electron Green's functions whereas dashed lines are impurities. $\hat C({\bf
		r}-{\bf r}';{\bf p},{\bf p})$ in Eq.~(\ref{eq:Z_ab}) is obtained by
	integration of this ladder over energy.}
\label{fig:cooperon}
\end{figure}

Although our conclusions are quite general, for simplicity we consider  the
situation  where the characteristic  radius of the grain is of order of the zero
temperature  superconducting coherence length and the value of the order
parameter in the puddles is much smaller than that in pure bulk superconductors,
$\Delta\ll \Delta_{0}$. This situation applies, for example, near the point
of a quantum superconductor-metal transition, where the typical distance
between the superconducting grains is larger than their size, which is of order
the zero temperature coherence length.~\cite{spivakZhouZyuzin} In this case, at
large separations between the grains, the Josephson coupling energy  can be
written in the form
\begin{equation}\label{eq:E_ab}
    E_{ab}^{\xi_a\xi_b}= 2\,  \mathrm{Re} \left[ e^{i (\phi_a -\phi_b)} Z_{ab}^{\xi_a\xi_b}\right],
\end{equation}
where $Z_{aa'}^{\xi\xi'}$  is given by
\begin{equation}\label{eq:Z_ab}
	Z_{aa'}^{\xi\xi'} =
	\mathrm{tr}  \int d {\bf r} d {\bf r}'d {\bf p} d{\bf p}'
		\hat\Delta_{a}^\xi({\bf r},\mathbf{p})\hat
		C(\mathbf{r}-\mathbf{r}'; {\bf p},{\bf p}')
		\hat\Delta_{a'}^{\xi'\dagger}({\bf r}',\mathbf{p}').
\end{equation}
Here $\mathrm{tr}$ denotes the trace over all  spin indices, and  $\hat
C(\mathbf{r}-\mathbf{r}'; {\bf p},{\bf p}')$ is the integral over energies of the Cooperon
diagrams  illustrated in Fig.~\ref{fig:cooperon}. 
The exchange energies $J_{aa'}^{\xi\xi'}$ and the 
phase $\theta_{aa'}(\xi,\xi')$ are related
to the modulus and phase of $2Z^{\xi\xi'}_{aa'}=J_{aa'}({\xi\xi'})\exp\left[i
\theta_{aa'}(\xi,\xi')\right]$.

\subsubsection{singlet pairing \label{sec:singlet}}

Let us begin by  considering the case where the Cooper pairing occurs in the
singlet channel $\hat\Delta^\xi_a=i\hat\sigma_2\Delta^\xi_a({\bf r},{\bf p})$,
which includes $s$ and $d$-wave superconductors.  In the presence of disorder,
even in the case where the clean bulk phase is a pure $d$-wave superconductor,
the order parameter in each grain will contain an $s$-wave component
\begin{equation}\label{eq:DeltaSD}
	\Delta^\xi_a({\bf r}, {\bf p}) =
	\Delta^{(s),\xi}_{a}(\mathbf{r},\mathbf{p}) +
	\Delta^{(d),\xi}_{a}(\mathbf{r},{\bf p}).
\end{equation}
where upper script in the parenthesis stands for the orbital symmetry whereas
$\xi$ 
indicates the Ising  variable which specifies which of the two time-reversed versions of the gap function is being considered.
Substituting Eq.~(\ref{eq:DeltaSD}) into Eq.~(\ref{eq:Z_ab}) and evaluating the
Cooperon we get three terms corresponding to $s$-$s$, $s$-$d$, and $d$-$d$
Josephson coupling
\begin{equation}\label{eq:Z_decomposition}
  Z^{\xi\xi'}_{aa'}=
	Z^{(ss),\xi\xi'}_{aa'}+
	Z^{(dd),\xi\xi'}_{aa'}+
	Z^{(sd),\xi\xi'}_{aa'}.
\end{equation}
At distances long compared to $p_F^{-1}$ but small compared to the thermal dephasing length the $s$-$s$  component is given by
 \begin{equation}\label{eq:Z_ss}
	 Z^{(ss),\xi\xi'}_{aa'}\propto \frac{\nu}{|{\bf r}_{a}-{\bf r}_{a'}|^D}\,
	 \langle\Delta^{\xi}_a \rangle \langle \Delta^{\xi'*}_{a'}
		 \rangle ,
\end{equation}
where $\nu$ is the density of states at the Fermi level, $D$ is dimensionality
of the system, ${\bf r}_a$ and ${\bf r}_{a'}$ are the locations of grains $a$
and $a'$, and $\langle\Delta^{\xi}_a\rangle$ denotes the order parameter
integrated over the single grain,
\begin{equation}\label{eq:Delta_average}
	\langle\Delta^{\xi}_a \rangle =
	\int  d \mathbf{r} d{\bf p} \Delta^{\xi}_a(\mathbf{r},{\bf p}) .
\end{equation}
Strictly speaking the slow power law decay of the Josephson coupling constant in
Eq.~(\ref{eq:Z_ss}) leads to a logarithmic divergence of the ground state
energy. However, multiple Andreev reflections~\cite{Andreev} of diffusing
electrons from the grains provide a cutoff of this divergence at large distances.\cite{spivakZhouZyuzin}
Since the cutoff length is greater
than the typical distance between the grains our results are not affected by the
presence of this cutoff.

In the same long-distance limit, the $s$-$d$  and  $d$-$d$ contributions are given by
\begin{equation}\label{eq:Z_sd}
Z^{(sd),\xi\xi'}_{aa'}\propto \nu\,
 \langle \Delta^{\xi *}_{a} \rangle Q^{
 \xi'}_{a',ij}\partial_i \partial_j
 \frac{1}{|{\bf r}_{a}-{\bf r}_{a'}|^D}  ,
\end{equation}
and
\begin{equation}\label{eq:Z_dd}
Z^{(dd),\xi\xi'}_{aa'}\propto \nu \,
 Q^\xi_{a,ij}Q^{\xi' \dagger}_{a',kl}\partial_i \partial_j \partial_k
 \partial_l\frac{1}{|{\bf r}_{a}-{\bf r}_{a'}|^D}.
\end{equation}
In the above formulas the $d$-wave component of the order parameter in grain $a$
is described by the second rank tensor $Q^\xi_{a,ij}$. For example for a spherical
Fermi surface in which $\Delta^{(d),\xi}_a({\bf r},{\bf p})=Q^\xi_{a,ij}({\bf r})p_i p_j$
(with $Q^\xi
_{a,ii}({\bf r})=0$)  we have
\begin{equation}\label{eq:Q_ij}
    Q^\xi_{a,ij}=\int d{\bf r} Q^\xi_{a,ij}({\bf r}).
\end{equation}

It is important to note that  $Z^{(sd)}_{ab}$ and $Z^{(dd)}_{ab}$ fall off faster with
the distance between the grains than $Z^{(ss)}_{ab}$. Therefore at large inter-grain
separations they can be neglected. The leading term $Z^{(ss)}_{ab}$ given by
Eq.~(\ref{eq:Z_ss}) has a phase factor that can be written as a 
sum of phase
factors of individual grains which are independent of the direction of the link
$\mathbf{r}_a-\mathbf{r}_b$. Therefore we arrive at the Mattis model,
Eqs.~(\ref{eq:JosEn}) and (\ref{eq:theta_Mattis}), where $\theta^\xi_a$ is the phase
of $\langle \Delta_a^\xi \rangle $ in Eq.~(\ref{eq:Delta_average}).  Indeed, in this limit, $J_{ab}(1)=J_{ab}(-1)\equiv J_{ab}$ is
independent of $\xi_a$ and $\xi_b$.

\subsubsection{triplet pairing \label{sec:triplet}}

Let us now turn to triplet superconductivity 
and to begin with in the case in which spin-orbit coupling is negligible.
Even in the case where $p_x+ip_y$ superconductivity occurs in the absence of disorder, in a particular grain
the order parameter will acquire an admixture of other $p$-wave components.
However,  the triplet and singlet
components of the order parameter do not mix.  In this case we get from
Eq.~(\ref{eq:Z_ab}) the following form of the Josephson coupling,
\begin{equation}\label{eq:Z_pp}
		Z^{(pp),\xi\xi'}_{aa'}\propto \nu \,  A^{\xi,\alpha}_{a,i}A^{\xi',\alpha
			*}_{a',j}\partial_i
		\partial_j \frac{1}{|{\bf r}_{a}-{\bf r}_{a'}|^D} ,
\end{equation}
where the matrix $A^{\xi,\alpha}_{a,i}$ describes the $p$-wave order parameter in grain $a$. For
example, for a spherical Fermi surface where
$\hat{\Delta}_a(\mathbf{r},\mathbf{p})=\hat\sigma_\alpha A^{\xi,\alpha}_{a,i} (\mathbf{r}) p_i$ it is given by
\begin{equation}\label{eq:A}
	A^{\xi,\alpha}_{a,i}= \int d \mathbf{r}
	A^{\xi,\alpha}_{a,i} (\mathbf{r}).
\end{equation}
The phase of the Josephson coupling in Eq.~(\ref{eq:Z_pp}) depends on the
relative orientation between the spatial structure of the order parameter
$A^{\xi,\alpha}_{a,i}$ (where the index $i$ indicates a preferred axis) and the direction of
the bond between the grains.  As a result the phase of $Z^{(pp),\xi\xi'}_{aa'}$  in
Eq.~(\ref{eq:Z_pp}) cannot be represented in the form of
Eq.~(\ref{eq:theta_Mattis}) in which the phases $\theta_a$ and $\theta_{a'}$ depend
only on the grain properties but not on the direction of the link connecting
them, ${\bf r}_a-{\bf r}_{a'}$. This means that we obtain a Josephson junction
array with frustration.

However, in the presence of spin-orbit interactions in non-uniform
superconductors the singlet, $\Delta$, and triplet, $\mathbf{\Delta}$, components
of the order parameter mix. In this case, at large separations between
grains, the Josephson coupling is again dominated by the $s$-wave component of the
order parameter and is described by Eq.~(\ref{eq:Z_ss}), which again leads us to
the Mattis model, Eqs.~(\ref{eq:JosEn}), (\ref{eq:theta_Mattis}).

\section{
Corrections to Mattis model and the global phase diagram}
\label{sec:corrections}

\begin{figure}[ptb]
\includegraphics[width=0.45\textwidth]{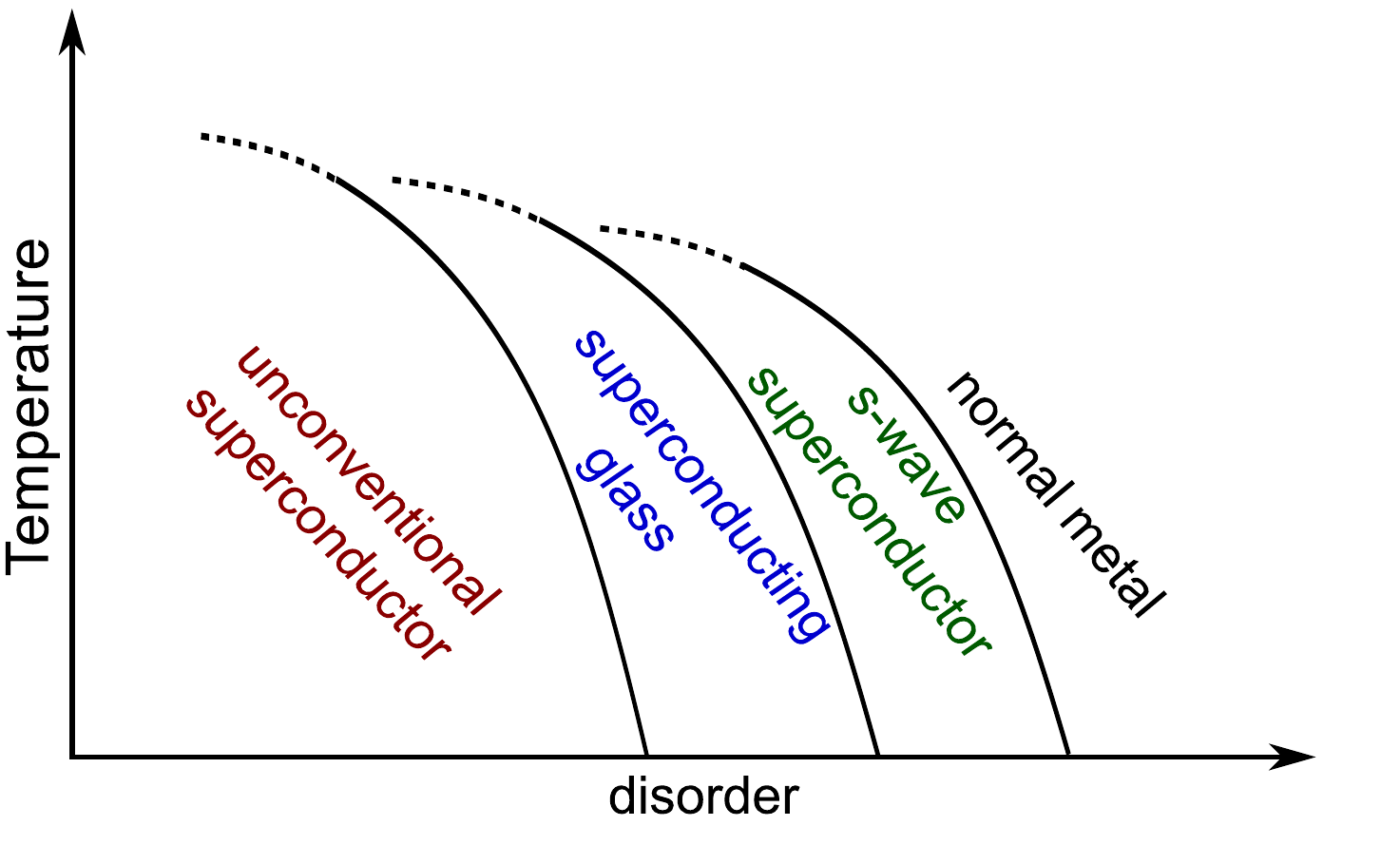}
\caption{Schematic picture of the phase diagram of an unconventional
	superconductor as a function of temperature and disorder}
\label{fig:phase_diagram}
\end{figure}

We have shown that at sufficiently strong disorder the properties of disordered
unconventional superconductors at long spatial scales may be described by the
effective Hamiltonian Eq.~(\ref{eq:JosEn}). To leading order in
inverse powers of the typical intergrain distance, $R$,
Eq.~(\ref{eq:JosEn}) to the Mattis model Eq.~(\ref{eq:JosEnMattRed}).
In this limit, the random phases of  the Josephson couplings between islands
can be gauged away;  as can seen from the construction that leads to Eq.~(\ref{eq:JosEnMattRed}),
the ordered state
corresponds to uniform-phase (ferromagnetic)  ordering of the s-wave
components from grain to grain,
even when the largest component of the order parameter on each grain is
unconventional.
Thus, at long distances the system behaves as an s-wave superconductor with respect to all
superconducting interference experiments.  For example, corner squid experiments
\cite{tsuei}  will not exhibit trapped fluxes. In the case where
in the absence of disorder there is
a $p_x\pm ip_y$ state, where the disorder is nearly strong enough to quench the superconductivity, there will be no edge currents,
and in the presence of an external magnetic field, the
long distance (topological) vortex structure will be that of a conventional s-wave superconductor.

Corrections to the Mattis model come from the coupling between the non-s-wave
components of the order parameter, and are thus smaller than the leading
contributions to $J$  in proportion to a positive power of $1/R$.  However, even
when $R$ is large, these corrections  can be qualitatively significant:
Since the $s$-$s$ contribution to
$J_{aa'}({\xi\xi'})$
from Eq.~(\ref{eq:Z_ss}) 
is independent of the pseudo-spin variables $\xi$
and $\xi'$,
to leading order in $1/R$  the energy of the system is
$2^N$-fold
degenerate, where $N$ is the number of grains.

The leading correction to the Josephson coupling energies  have the
form of
either Eq.~(\ref{eq:Z_sd}) or (\ref{eq:Z_pp}).
In the presence of these corrections,  the
energy depends on the configuration of the pseudo-spins and
there is a level of frustration which cannot be removed (as in the Mattis model) by a gauge transformation.  Specifically,
although in the expression for the Josephson energy,
$\sum_{ab}
\tilde
J_{ab}(\xi_a\xi_b)\cos(\tilde{\phi}_{a}-\tilde{\phi}_{b}+\tilde{\theta}_{ab}^{\xi_a\xi_b})$,
$\tilde{\theta}_{ab}$ are small,
\[
\tilde\theta^{\xi_a\xi_a}_{ab}=\mathrm{Im} \ln
\frac{Z^{(pp),\xi_a\xi_b}_{ab}}{Z^{(ss),\xi_a\xi_b}_{ab}} \ll 1,
\]
they reflect intrinsic frustration in the couplings since the sum of the phases around a typical closed loop, $\sum_\bigcirc \tilde\theta$, is generally non-zero.
Moreover, both  $\tilde J_{ab}$ and $\tilde \theta_{ab}$ depend on $\xi_a$ and $\xi_b$.

Therefore, one generic
consequence of the corrections to the Mattis model is that they  lift  the
energy degeneracy of the system with respect to the pseudo-spin variables $\xi$.
Consequently, one expects the subsystem of pseudo-spins to form a glassy state.  Another consequence of
the corrections is that they result in the existence of equilibrium currents.
In the three dimensional case, the existence of the  corrections to the Mattis model
does not destroy the long range superconducting order
characterized by the
phase $\tilde{\phi}$. Therefore,
results from the Mattis model concerning the long-range  s-wave-like nature of
the superconducting
state are robust to these corrections. In two spatial
dimensions the correction terms
necessarily eliminate long range
phase coherence, since the
correlation function of the phases in the ground state diverges logarithmically
at large distances.  However, as long as $\tilde\theta^{\xi_a\xi_b}_{ab} \ll 1 $, the
length at which the phase changes by a number of order unity is exponentially
large  in comparison to the intergrain distance.

At intermediate strength of disorder,  when
$R$ gets to be comparable to the size of the superconducting islands,  the effective energy of the
system,  Eq.~(\ref{eq:JosEn}), cannot even approximately be reduced to the Mattis model,
Eq.~(\ref{eq:JosEnMattRed}). The phases  $\tilde\theta_{ab}$ that cause
frustration are then of order unity. In this case the system is a superconducting
glass.\cite{Kivelson,Kivelson1}. The global phase diagram of the system  is
schematically shown in Fig. \ref{fig:phase_diagram}.

\acknowledgments   Work at Stanford University was supported by DOE Office of Basic Energy Sciences DE-AC02-76SF00515. Work at the
University of Washington was supported by U. S. Department of Energy
Office of Science, Basic Energy Sciences under award number
DE-FG02-07ER46452. B.S. thanks the International Institute of
Physics ( Natal, Brazil) for hospitality during the completion of the paper.


\begin{thebibliography}{99}
\bibitem{caveat} There are, however, some subset of experimental observations that are not easily reconciled with the existence of a chiral p-wave state in Sr$_2$RuO$_4$.  See, for example, Refs. \onlinecite{moler} and \onlinecite{lederer} for a discussion.
\end{thebibliography}
\end{document}